\documentclass{sig-alternate}

\usepackage[
  pass,% keep layout unchanged 
]{geometry}

\usepackage{graphicx}
\usepackage{paralist}
\usepackage{todonotes}
\usepackage{array}
\usepackage{multirow}
\usepackage[nolist]{acronym}
\usepackage{enumitem}
\usepackage[utf8]{inputenc} %added by bjorn with two dots :)
\newcommand{\specialcell}[2][c]{%
  \begin{tabular}[#1]{@{}c@{}}#2\end{tabular}}

\begin{document}
\toappear{}
%
% --- Author Metadata here ---
\conferenceinfo{First International Workshop of Open Innovation in Software Engineering}{'15 Tallin, Estonia}
%\CopyrightYear{2007} % Allows default copyright year (20XX) to be over-ridden - IF NEED BE.
%\crdata{0-12345-67-8/90/01}  % Allows default copyright data (0-89791-88-6/97/05) to be over-ridden - IF NEED BE.
% --- End of Author Metadata ---

%\title{Researching Requirements Engineering in Open Innovation}
\title{Requirements Engineering in Open Innovation: \\ A Research Agenda}


\numberofauthors{3} %  in this sample file, there are a *total*
\author{
% 1st. author
\alignauthor
Johan Lin{\aa}ker\\
       \affaddr{Lund University, Sweden}\\
       \email{Johan.Linaker@cs.lth.se}
% 2nd. author
\alignauthor
Bj\"{o}rn Regnell\\
       \affaddr{Lund University, Sweden}\\
       \email{Bjorn.Regnell@cs.lth.se}
\alignauthor
Hussan Munir\\
       \affaddr{Lund University, Sweden}\\
       \email{Hussan.Munir@cs.lth.se}
}

\maketitle
\begin{abstract}
In recent years Open Innovation (OI) has gained much attention and made firms aware that they need to consider the open environment surrounding them. To facilitate this shift Requirements Engineering (RE) needs to be adapted in order to manage the increase and complexity of new requirements sources as well as networks of stakeholders. In response we build on and advance an earlier proposed software engineering framework for fostering OI, focusing on stakeholder management, when to open up, and prioritization and release planning. Literature in open source RE is contrasted against recent findings of OI in software engineering to establish a current view of the area. Based on the synthesized findings we propose a research agenda within the areas under focus, along with a framing-model to help researchers frame and break down their research questions to consider the different angles implied by the OI model.
\end{abstract}



\section{Introduction}

Due to the eruption of new ways in communicating and working distributed, along with e.g. lowered barriers for startups and increased access to knowledge sources, software producing firms now need to explore and exploit their outside surroundings to a higher degree than before to complement their internal innovation capabilities~\cite{chesbrough2003open}. This phenomenon is captured by the paradigm of Open Innovation (OI), defined by Chesbrough~\cite{chesbrough2003open} as \textit{``a paradigm that assumes that firms can and should use external ideas as well as internal ideas, and internal and external paths to market, as they look to advance their technology''}. The OI model incorporates multiple applications. In Software Engineering (SE) the most common example probably is Open Source Software (OSS)~\cite{munir2014open}, which has been around for decades and proven to be a useful and important building block in product strategies among software producing firms. The emergence of the OI paradigm has highlighted this fact even more in the sense that OSS may be used to better leverage and advance firms' internal technical and innovation capital~\cite{west2003open}.
%Power of the crowd is a pertinent concept that relates very well to OSS, but also to crowdsourcing which also falls under the OI definition. It makes up a central opportunity as well a challenge with which SE practices has yet to catch up, in order to fully accommodate and maximize the innovation outcome that firms aspire to achieve~\cite{wnuk2013engineering}. 
Requirements Engineering (RE), which is pivotal for a software development project to succeed, needs to take the changes implied by OI in regard and adapt accordingly, e.g. to consider how firms should reach out, explore and manage the new potential stakeholders, users and customers, for which requirements can be elicited from or outsourced to.

In this paper we build on and advance the proposed SE framework for fostering OI by Wnuk and Runeson~\cite{wnuk2013engineering}. We also propose a research agenda with several research questions regarding RE practices in relation to OI and how these could or should be adapted in order to enable for and optimize the outcome for firms using an OI strategy. The proposed research agenda suggests a number of research directions in different sub-disciplines of RE. The research questions elaborated on in this paper are in line with the proposal by Wnuk et al.~\cite{wnuk2013engineering} to \textit{``\ldots focus on release-planning, stakeholder analysis, trade-off between effort (cost) and value and the degree of innovation in candidate features needed in evolving systems, to take significant future market shares in open innovation software development''}, and our main conjecture is that these trade-offs constitute essential and hard problems of future software RE, and these problems are further complicated in an OI context. 

In section two, the methodology is presented in short. In section three, synthesized results from OSS RE and OI literature are presented. In section four a model is presented to help frame RE in an OSS and OI setting along with suggestions for future research. The paper is concluded in section five.

\section{Methodology}
Based on the research direction within RE in OI suggested by Wnuk et al.~\cite{wnuk2013engineering}, we conducted a literature survey of RE practices in OSS within the areas of stakeholder identification and management, contribution strategies and opening up criteria, and prioritization and release planning. These results were compared and contrasted against recent findings of OI in SE reported by Munir et al.~\cite{munir2014open}. Findings where synthesized in the categories Stakeholder management, When to open up, and Prioritization and release planning. Building further, we constructed a research agenda within the areas under focus, along with a framing-model to help researchers frame and break down their research questions to consider the different angles implied by the OI model.

\section{Contrasting Literature}
On the basis of our literature survey, we elaborate and contrast between RE in OI and OSS in regards to the areas of Stakeholder management, When to open up a firms software, and Prioritization and Release planning.

\subsection{Stakeholder Management}
We refer to the definition of Glinz and Wieringa~\cite{glinz2007guest} and define a stakeholder as \textit{``\dots a person or organization who influences a system's requirements or who is impacted by that system''}, with system meaning an OSS project. With stakeholder management we consider both identification of the stakeholders and analysis of their interest or intent in an OSS project and its community. This is a problematic area due to that the open environment implied by OI allows for an influx of new and earlier unknown stakeholders into a firm's formerly closed domain. Staying aware of who else are present in a community and what their intents are may therefore prove difficult. Ignorance may constitute a risk and jeopardize a firm's stake in the OSS. Also, with an awareness, opportunities could arise as stakeholders with aligning intents and complementary competences may be found. Another issue besides awareness, is managing conflicts and different needs of the stakeholders~\cite{munir2014open}. As they may have clashing intents with business stakes involved, creating a common vision becomes troublesome~\cite{dahlander2008firms}. In such cases it may be beneficial to know who are aligned and those to persuade, as some stakeholders could be unknown and reside silent until confronted. 

Stakeholder management may hence be considered as an important area to master for a firm to adapt itself to an OSS community, as well as to gain and maintain a profitable position with sustainable influence in the community's governance structure~\cite{dahlander2008firms}. Consequently, it may further be considered an enabling factor for firms to more effectively be able to perform other RE practices such as elicitation, prioritization and release planning in relation to the community~\cite{glinz2007guest}. 

In an OSS community stakeholders may be identified by the natural means they communicate, both electronically (e.g. mailing lists, forums, IRC) and in-person (e.g. conferences, hackathons). For the former type, a multitude of quantitative analytical methods exists to identify present stakeholders (e.g.~\cite{lim2011stakesource2, laurent2009lessons}). Tools also exist, which allow for further analysis to be performed on the earlier mentioned software repositories, but also on other types, e.g. source code repositories and issue trackers~\cite{robles2009tools}. %Due to the variability of what stakeholders are present and what level of influence they have, it would be interesting to analyze these change patterns as it can help a firm to adapt and find new partners.
However, quantitative analysis can only bring you to a certain point. To get a clearer view, a qualitative analysis could be performed as well, either by observing, engaging or a combination of them both. Regards should be taken to both the electronic and in-person communication channels. As social events and live meetings are becoming increasingly popular, firms may need to consider distributing their presence offline as well~\cite{stam2009does}. Some stakeholders may not be as vocal online as they are offline, why these could risk going unnoticed. Furthermore, by showing one self and giving notice on one's intents and knowledge, this may attract some stakeholders to come out of the shadows. 

Different kinds of stakeholders may be found in different communication channels, 
%why it may be needed to cover as broad and diverse ground as possible. On the other hand, if one knows what type of stakeholder is of interest this could help limit the search. 
Knauss et al.~\cite{knauss2014openness} describe how requirements are communicated between stakeholders in two different flows in an open-commercial software ecosystem, which also transfers to an OSS community. First, an emergent requirements flow captures requirements as developers and users find new ways of using the software and present these on a work-item level, much as how e.g. prototypes and other informalisms are used as post-hoc assertions in OSS RE~\cite{scacchi2002}. Secondly, a strategic flow is used to communicate the business goals and strategies of the different stakeholders. %Which communication channels, how they are used within these flows and how they may differ between communities needs further exploration. 

For analyzing the stakeholders' intents similar methodologies may be applied as for identification. Regarding quantitative methods, Gonzalez-Barahona et al.~\cite{gonzalez2013understanding} describes how patterns may be found in how other stakeholders contribute (e.g. feature implementations, bug-fixes, discussions), how the projects generals stance is on certain topics, and how stakeholders work together (e.g. on implementation, decision-making). This information may then be used to analyze their intents and strategies (e.g. what functionality they value most, in what direction they are striving towards, how committed they are), and to find potential collaboration opportunities. Modeling the relationships between the stakeholders may be a further step to help create a better picture. Fricker~\cite{fricker2010requirements} suggests how relations amongst stakeholders in software ecosystems may be modeled as requirements value chains, whilst Damien et al.~\cite{damian2007collaboration} suggests using social network analysis for stakeholders in distributed development. 

%As Knauss et al.~\cite{knauss2014openness} suggests, future research should be done in the lines on how to manage stakeholders in an open environment, as well as how the single firm should use the open requirements flows to position themselves in relation to the other stakeholders, specifically with regards to the context of OI. 

%\todo[inline]{Include proposed research questions explicitly?? I think no \\ -- BR Thinks YES! :)}
%This leads us to define the following research questions:

\subsection{When to Open Up}
Reasons for opening up and releasing a software project, or parts of it, could be many, and often of a strategic nature~\cite{dahlander2008firms}. On an upper management level, such decisions may be rare, and with time to analyze. In cases where firms do continuous development and exploitation of OSS, the question can be more common. Such lower level decisions may not have the time needed to be processed through traditional flows in large software producing firms. The pace of RE may vary between OSS communities, but in instances where the RE process is of a just-in-time character decisions may need to made quick. For firms having a maintainer's role, feature requests from the OSS community could be reoccurring and developers and lower level managers need to know when they should implement such requests, reject or leave be for others in the community to act on. In the role of a contributor, this includes when to implement certain feature requests posted in the community, as well as if and when they should contribute their internal feature implementations. A feature's differentiating value needs to be weighed against the cost of maintaining it internally, e.g. though patching each release of the OSS~\cite{wnuk2012can, ven2008challenges}. Structuring such strategies would help developers prioritize and spend resources on implementing features which has the highest business value to the firm~\cite{wnuk2012can}. If not, this could be a notable cost, but also a strategic loss if differentiating parts are wrongly revealed.
%In a large setting this would risk being an expensive matter and a notable lost opportunity cost, if developers are allowed to implement non-critical or relevant bugs and features freely.

Through selective revealing~\cite{henkel2014emergence}, certain parts of the code could be broken out and contributed. Separating the parts of differentiating value may however still prove difficult~\cite{wnuk2012can}. One way of tackling this issue would be to provide certain parts as enablers, while the innovative features are kept internal~\cite{west2003open}. Another strategy could be to disclose the technology, but under such circumstances that it will not be of value for competitors, e.g. through licensing~\cite{west2003open}. Related questions include when in the product and technology life-cycle this should be done~\cite{van2009commodification}, and how to make it happen practically as the packaging of the selected part or feature must be accepted by the community and integrate well with the architecture and other functionality. Often features are analyzed and judged by the community based on a prototype or finished implementation~\cite{scacchi2002}. This again could be communicated in several ways, e.g. electronically or live at social community gatherings~\cite{stam2009does}. Certain situations may require pro-active judgment and contribution. In an example where a firm wants a community to head in a certain direction and add support for a certain functionality or environment~\cite{dahlander2008firms}, the right way to go could be to incrementally contribute features building up to the end-goal. Again comes the question on how this should be done practically. 

Hence, such practices and guidelines on how and when firms should contribute or receive contributions need to be integrated into the RE processes in order to streamline development and keep it aligned with the firms business goals and long-term strategy. Thereby, selective revealing or contribution strategies~\cite{munir2014open, wnuk2012can, ven2008challenges} may link together the strategic intents of the firm with its lower level RE processes to better exploit its engagement with the OSS community and maximize the innovation outcome.

%This leads us to define the following research questions:

\subsection{Prioritization and Release Planning}
Prioritization is an RE practice where the selection of requirements for a certain release is decided. Release planning is about packaging the right features into the right product, in time with as high satisfaction among the stakeholders as possible. In OI this area has had limited coverage with some exceptions~\cite{wnuk2012can, nayebi2014open} compared to the field of OSS (e.g.~\cite{michlmayr2015and, wright2009subversion, michlmayr2007release, erenkrantz2003release}). Michlmayr et al.~\cite{michlmayr2007release} describes how release strategies used by OSS communities can be classified into feature-based and time-based. The former performs a release when a set goals have been fulfilled, e.g. a certain set of features has been implemented. The latter performs releases according to a preset schedule. Hybrid versions has also been reported~\cite{wright2009subversion}. Literature has however favored the time-based strategy~\cite{michlmayr2015and,wright2009subversion}. With fixed release dates and intervals, firms can better adapt their internal plans so that additional patchwork and differentiating features may be added in time for product shipment to market~\cite{michlmayr2015and}. Other issues reported by Michlmayr et al.~\cite{michlmayr2015and}, which could be refrained with a time-based release strategy, are rushed implementations, workload accumulation, outdated software and delayed releases. 

In regards to OI, Nayebi et al.~\cite{nayebi2014open} proposes a decision support methodology called Analytical Open Innovation, which considers information gathered through crowdsourcing from customers and stakeholders, and analyzes their expectations. This information can help adapt product composition and release planning to predicted trends and anticipated user expectations. Wnuk et al.~\cite{wnuk2012can} reports on how release planning is conceived as more challenging in an OI context. One reason is that innovative features included in the normal requirements flow are killed because they do not meet deadlines and sharp acceptance criteria imposed by the market. A separate flow for innovative requirements is proposed so that innovative requirements are able to mature in their own pace. Once mature enough they would then be able to enter the normal and more ''harsher'' requirements flow. Further reason for release planning being extra complicated is due to the lack of control over the process used in the OSS communities~\cite{wnuk2012can}. An example describes how it would have been a better decision do adopt the OSS and sell it with minimum changes, rather than have spent resources on adding differentiating features, which is a common challenge in OSS release planning~\cite{michlmayr2015and}. This could imply a need for firms to better understand the release processes used by OSS communities, as well as to propagate for more structured and predictable processes. 

%From an OI perspective it would be interesting to further investigate what release strategy is optimal under what conditions, how practices and tools should be adopted~\cite{michlmayr2007release}, how care should be taken to involvement of other corporate stakeholders~\cite{michlmayr2015and}, how risk and dependency issues should be considered~\cite{wnuk2012can}, but also how to tailor internal release processes to fit with the OSS community's in regards to innovative requirements. Specifically, how the separate requirements flow suggested by Wnuk et al.~\cite{wnuk2012can} would fit and be structured in such a context. Even for cases where a firm is not a core team member of a community, it is important to know how an ideal process would look like and how they would be able to promote such an implementation in a community~\cite{michlmayr2015and}. 

Utilizing the overflow of information and access to new data sources constitutes an opportunity to be exploited, as exemplified by the analytical approach suggested by Nayebi et al.~\cite{nayebi2014open}. The factor of a requirements innovativeness and how this should be regarded in the prioritization and release planning processes is another example where this could be profitable. Establishing metrics that regards a feature's innovation capability, output and performance~\cite{edison2013towards} would be helpful inputs to optimize the innovation outcome through the prioritization and release planning processes. There are frameworks that considers innovativeness in the release planning process (e.g.~\cite{regnell2012requirements, khurum2009innovative}) but the area still lacks thoroughly investigated and validated models, especially in regards to the OI context.

\section{Future Research}
Based on the contrasting between RE in OSS and OI in the previous section, we formulate suggestions for future research directions. To further frame and break down questions suggested, and help researchers consider the many angles implied OI, we begin by defining a framing-model.

The OI model is often illustrated by a funnel~\cite{chesbrough2003open}, here presented in Figure 1. The funnel itself represents the firm internally, which in a SE context we define as the software development process. This may in turn be illustrated either as an iterative process, or just as one single development cycle (e.g. from RE, design, implementation, test, to release). The funnel is full of holes representing openings between the firm's development process and the open environment. The arrows going in and out represent transactions and exchange of knowledge between the two. This may for example constitute of feature requests or bug reports, design proposals or feature road maps, feature implementations or bug fixes, test cases, complete sets of code as in plugins, components, platforms or even products. From a more specific RE perspective this includes activities such as elicitation, prioritization and release planning, which are performed interactively between the internal and the open environment.

%As illustrated, the exchange of this knowledge may be bi-directional, i.e. it can go into the development process from the open environment (\textit{outside-in}), or from the development process out to the open environment (\textit{inside-out}). On the left-hand side the arrows represent the intent and business model, which encompasses and motivates the development process. On the right-hand side the arrows represent the output from the development process, hopefully with some degree of innovativeness. This output may be a new or improved product or service from which the company may capture or create value with the help of their business model. Also on the right-hand side are the bulls-eye marks, which represents the current but also new alternative markets to which the firm's business model will deliver the product or service produced from their development process.

%If we reconnect with the definition of OI, the transactions moving in and out of the funnel represents the use of external and internal ideas, whilst the arrows coming out on the right-hand side represents the internal and external paths to market, all with the overall goal to create and capture value, and to advance the firms technology.

\begin{figure}[hbtp]
\vspace{0pt}
\centering
\scalebox{1}{
\includegraphics[scale=.33]{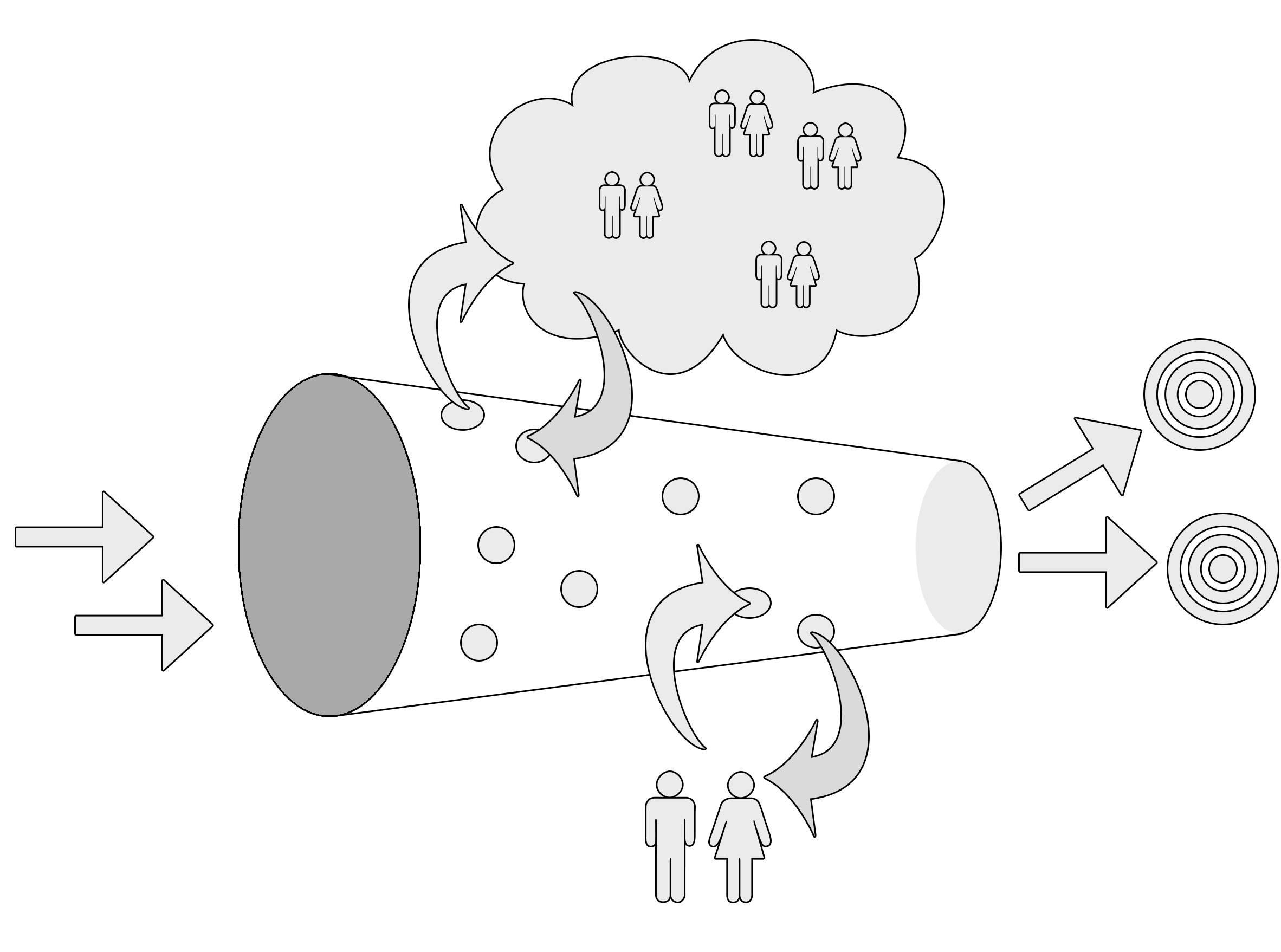}
}
\label{fig:Funnel}
\caption{The OI model illustrated with interactions between the firm (funnel) and its external collaborative actors. Adopted from Chesbrough~\cite{chesbrough2003open}.}
\vspace{-16pt}
\end{figure}

Firstly, based on Figure 1, we make a distinction between an internal and external RE process. The internal concerns the normal process connected to a firm's internal product development. The external concerns the one used in an OSS community with which the firm interacts. This distinction makes it clearer to illustrate how e.g. an internal release process have to consider the external release process of an OSS community, how the external stakeholder management may differ from the internal, or how parts of the internal software may be revealed to an existing or new OSS community, as well as sourced externally from an OSS community.

A second distinction can be made between practices inferred from the OI and OSS RE literature respectively. In OSS RE literature, the scope has mainly been on a technical level with a starting points from seminal reports by e.g. Scacchi~\cite{scacchi2002} and Erenkrantz~\cite{erenkrantz2003release}. In the OI literature, RE has had limited coverage with a few notable exceptions focused mainly on release planning~\cite{wnuk2012can, nayebi2014open}. Focus has instead been on how firms on a strategic level may influence or connect itself to OSS communities, and how OSS should be leveraged from a strategic point of view and integrated in different kinds of business models~\cite{west2003open, munir2014open}. Similar reports on the governance structures, how these function and should be managed are also reported in the gray area between OSS and OI. This distinction opens up for an interpretation of RE in OSS and OI on two levels, a strategic and an operational. 

Finally, a third distinction can be made in regards to which direction the internal and external RE processes interact. A firm may e.g. reveal or outsource a feature to the OSS community (\textit{inside-out process}), acquire a feature from the OSS community (\textit{outside-in process}), or collaborate jointly with the OSS community on a feature (\textit{coupled process})~\cite{chesbrough2003open}.

These three distinctions are presented together in Table 1. Although all aspects of the framing-model may not be applicable in each case, it offers a tool to help angle and break down research questions to a possibly more concrete level. E.g. how a firm should adapt its release process towards an OSS community may be angled from the internal release process and the external release process used by the OSS community. Broken down further in regards to direction, we can separate between how and when a firm should adopt a new release, or contribute a feature to get it included in a certain release. And furthermore, we can distinct between how the firm should align these processes with their business goals and communicate on a strategic level, against technically managing dependencies and risks of features internally and externally on an operational level.

\begin{table}[h]
\vspace{-8pt}
\caption{Conceptualization of RE in the field of OSS and OI.}
\label{tbl:Model}
\begin{tabular}{|p{2.4cm}|p{1.5cm}|p{1.3cm}|p{1.5cm}|}
\hline
                    &\specialcell{Internal\\RE process}     &  \specialcell{Direction} &\specialcell{External\\RE process} \\ \hline
Strategic level     &                     &                 &  \\ \hline
Operational level   &                     &                 &  \\ \hline
\end{tabular}
\end{table}

The suggested research questions below should be further viewed from the framing-model presented in Table 1.

\subsection{Stakeholder Management}
\begin{enumerate}[leftmargin=.5in, label=\bfseries RQ A\arabic*]
\item \textit{How to identify new and stay aware of present stakeholders in an open environment (e.g.~\cite{lim2011stakesource2, laurent2009lessons, stam2009does})?}

\item \textit{How to continuously prioritize and judge importance of the stakeholders in an open environment (e.g.~\cite{gonzalez2013understanding})?}

\item \textit{How to manage, adapt and act in shifting governance structures in an open environment with multiple stakeholders and fluctuating partnership types (e.g. feature-by-feature, project, and product)?}

\item \textit{How to leverage the requirements flows to position oneself strategically in an open environment~\cite{knauss2014openness}?}
\end{enumerate}

\subsection{When to Open Up}
\begin{enumerate}[leftmargin=.5in, label=\bfseries RQ B\arabic*]
\item \textit{How to determine what artifacts (e.g. ideas, spill-over requirements, IP, plug-ins, products) to open up and to what degree (e.g. selective revealing~\cite{henkel2014emergence})?}

\item \textit{How to determine when the right moment is to open up the artifact for external involvement~\cite{van2009commodification}?}

\item \textit{How to determine the way in which an artifact is to be developed (e.g. co-develop, outsource) and with/by whom (e.g. single partners, groups or ecosystems)~\cite{van2009commodification}?}
\end{enumerate}

\subsection{Prioritization and Release Planning}
\begin{enumerate}[leftmargin=.5in, label=\bfseries RQ C\arabic*]
\item \textit{What release strategy is optimal under what conditions, and how should related practices and tools be adopted~\cite{michlmayr2007release}?}

\item \textit{How should care be taken to involvement of other corporate stakeholders~\cite{michlmayr2015and}?}

\item \textit{How should the internal release processes be tailored to fit with the OSS community's in regards to innovative requirements? Specifically, how would a separate requirements flow fit and be structured in such a context~\cite{wnuk2012can}?}

\item \textit{How should risks and dependencies of features be managed and considered in an open environment~\cite{wnuk2012can}?}

\item \textit{How should innovativeness be considered as a decision factor (e.g.~\cite{regnell2012requirements, khurum2009innovative})?}

\item \textit{What other decision factors may be considered relevant?}
\end{enumerate}

\section{Conclusions}
OI makes up a central opportunity as well as a challenge with which SE practices has yet to catch up, in order to fully accommodate and maximize the innovation outcome that firms aspire to achieve~\cite{wnuk2013engineering}. This includes RE, which is pivotal for a software development project to succeed. In response we build on and advance the proposed SE framework for fostering OI by Wnuk and Runeson~\cite{wnuk2013engineering}, focusing on stakeholder management, when to open up, and prioritization and release planning. After performing a literature survey on OSS RE, we contrast it against recent finding of OI in SE~\cite{munir2014open} to establish a current view of the area. Based on the synthesized findings we propose a research agenda along with framing-model to help researchers frame and break down their research questions to consider the different angles implied by the OI model. 

We acknowledge that the literature survey may not give a complete view of the current state of RE in OSS and OI. However, we believe it can still explain current practices to a valuable extent and provide a valid basis for our proposed research agenda. Further, the proposed framing-model and research agenda are not to be seen as exhaustive but rather show suggested directions and help researchers break new territory in the field of RE in OI and OSS. 

%\section*{Acknowledgment}
%This work is funded by the Swedish National Science Foundation Framework Grant for Strategic Research in Information and Communication Technology, project Synergies (Synthesis of a Software Engineering Framework for Open Innovation through Empirical Research), grant 621-2012-5354.

%\bibliographystyle{sig-alternate}
\bibliographystyle{abbrv}
\bibliography{refs.bib}

\end{document}